\documentclass[aps,twocolumn,showpacs,final,floatfix]{revtex4-1}
\usepackage[pdftex]{graphicx}  % needed for figures
\usepackage[update]{epstopdf}
\usepackage{amssymb}
\usepackage{amsmath}
\usepackage{color}
\usepackage{textcomp}
\usepackage[utf8]{inputenc}

\graphicspath {{./figures/}}
\makeatletter
\def\input@path{{./figures/}}
\makeatother

\begin{document}

\title{Quasi-exact ground-state algorithm for the random-field Potts model}

\author{Manoj Kumar}
\email[]{manojkmr8788@gmail.com}

\affiliation{Institut für Physik, Technische Universität Chemnitz, 09107 Chemnitz, Germany}

\author{Martin Weigel}
\email[]{martin.weigel@physik.tu-chemnitz.de}
\affiliation{Institut für Physik, Technische Universität Chemnitz, 09107 Chemnitz, Germany}
\date{\today}

\begin{abstract}
  The use of combinatorial optimization algorithms has contributed
  substantially to the major progress that has occurred in recent
  years in the understanding of the physics of disordered systems,
  such as the random-field Ising model. While for this system exact
  ground states can be computed efficiently in polynomial time, the
  related random-field Potts model is {\em NP\/} hard
  computationally. While thus exact ground states cannot be computed
  for large systems in this case, approximation schemes based on graph
  cuts and related techniques can be used. Here we show how a
  combination of such methods with repeated runs allows for a
  systematic extrapolation of relevant system properties to the ground
  state. The method is benchmarked on a special class of disorder
  samples for which exact ground states are available.
\end{abstract}
\maketitle

\section{Introduction}

Impurities are omnipresent in samples in the laboratory. Their theoretical
description in terms of quenched random disorder in magnetic systems represented by
spin models turns out to be an extremely challenging task that has attracted an
extensive amount of research activity in past decades \cite{young:book}. Disorder
has profound effects on the type of ordering and the nature of the associated phase transitions.
Much of the progress achieved to date towards an understanding of such systems has
been due to large-scale numerical simulation efforts. Standard approaches such as
canonical Monte Carlo simulations are heavily affected by the complex free-energy
landscapes composed of a multitude of metastable states separated by barriers that
are the signature of such systems \cite{janke:07}. More sophisticated techniques in
the form of generalized-ensemble simulations such as parallel tempering
\cite{geyer:91,hukushima:96a}, multicanonical simulations \cite{berg:92b,janke:03a,gross:17}
or, most recently, population annealing
\cite{hukushima:03,machta:10a,wang:15a,barash:16,weigel:21} lead to dramatically improved
performance in such situations \cite{kumar:20}, but they are not able to fully remove the slowing
down of dynamics induced by the combination of disorder and frustration.

For the case of random-field systems, where the renormalization group indicates that
the fixed point relevant for the critical behavior sits at zero temperature
\cite{bray:85a}, an alternative approach of analysis is based on the study of the
ground states of individual disorder samples. To arrive at such configurations one
might employ generic optimization methods such as simulated annealing
\cite{kirkpatrick:83} or genetic algorithms \cite{sutton:94,weigel:05f,weigel:06b} that provide
capabilities to overcome the inherent energy barriers and/or explore different
valleys independently, but such techniques do not constitute a magic bullet for
handling the complexity of the energy landscape. As was noted early on
\cite{dauriac:85}, for the random-field problem with Ising symmetry (RFIM) things are
somewhat easier in that the ground-state computation can be mapped onto a
maximum-flow problem for which efficient (polynomial-time) algorithms are
available~\cite{mainou:22}. This has enabled high-precision analyses of the critical behavior of this
model, see, e.g.,
Refs.~\cite{middleton:02,ahrens:11a,stevenson:11,fytas:13,fytas:16}.

For related, somewhat richer systems such as the random-field Potts model (RFPM),
however, the situation is less fortunate as the ground-state problem for more than
two spin states corresponds to optimizing a multi-terminal flow, a task that can be
shown to be {\em NP\/} hard \cite{boykov:01}. While it is hence not possible for this system to find
exact ground states in polynomial time, we have shown recently that
good approximations can be computed with reasonable time investment employing
suitable generalizations of the graph-cut (GC) methods used for the RFIM
\cite{kumar:18}. Algorithms for this purpose have previously been discussed in the
context of computer vision \cite{kolmogorov:04a}. In the following, we investigate
how a randomization of this approach allows to construct an extension that
systematically converges to the exact ground state. By constructing a particular set
of disorder samples for which exact results are available from a different approach
(TRW-S as proposed in Ref.~\cite{kolmogorov:06}), we study how the minimum energies
as well as state overlaps of the randomized method approach the exact result, thus
developing a technique for systematic extrapolation of the approximate data.

The remainder of this paper is organized as follows. In Sec.~\ref{model} we define the random-field Potts model in the variant discussed here and describe the graph-cut technique for computing approximate ground states. We then discuss how $n$ repeated runs with different initial conditions are used for a systematic improvement of results. This leads to $n$-dependent estimates of the thermodynamic quantities that are later used for extrapolation. In addition, we introduce the TRW-S method that allows us to generate a set of samples with the associated exact ground states. In Sec.~\ref{result} we report on our results for the extrapolation of approximate ground states. A detailed analysis of exact samples reveals that typical quantities approach their ground-state values in a double power-law fashion that is also shown to apply to the case of regular samples. This setup enables a reliable extrapolation of data for moderate values of $n$ to the $n\to\infty$ limit. Finally, Sec.~\ref{conclude} contains our conclusions.

\section{Model and methodology}
\label{model}

\subsection{The random-field Potts model and graph cuts}

The $q$-state RFPM considered here is governed by the Hamiltonian~\cite{blankschtein:84}
\begin{equation}
  \label{hamilt}
  \mathcal{H}=-J\sum_{\left<ij\right>}\delta_{s_i,s_j}-\sum_i\sum_{\alpha=0}^{q-1}h_{i}^{\alpha}\delta_{s_i,\alpha},
\end{equation}
where $\delta_{x,y}$ is the Kronecker delta function. According to the Potts
symmetry, the spins $s_i$ take values from the set $\{0,1,....,q-1\}$. The variables
$\{h_{i}^{\alpha}\}$ are the quenched random fields at site $i$, acting on state
$\alpha$, and each is drawn independently from a normal distribution,
\begin{equation}
  P(h_i^\alpha) = \frac{1}{\sqrt{2\pi}\Delta} \exp \left[-\frac{({h_i^\alpha})^2}{2\Delta^2} \right].
  \label{eq:ditrib}
\end{equation}
The variance $\Delta$ determines the strength of disorder. Different ways of exposing
the Potts spins to random fields have also been considered
\cite{goldschmidt:85,eichhorn:95}, especially for the case of discrete random-field
distributions. While we did not consider such variations explicitly, we expect the
general results discussed in the present study to carry over to such generalized
disorder distributions. 

For $q=2$, it can be easily seen that the RFPM Hamiltonian
in Eq.~(\ref{hamilt}) corresponds to the RFIM. In this case, ${\cal H}$ can be written as \cite{kumar:18},
\begin{equation}
\begin{split}
  \mathcal{H} = & -\frac{J}{2}\sum_{\left<ij\right>}[\sigma_i\sigma_j+1] \\
  & -\frac{1}{2}\sum_i[(h_i^+-h_i^-)\sigma_i+(h_i^++h_i^-)],
\end{split}
\end{equation}
where $h_i^+$ and $h_i^-$ represent the two field components with $\alpha=\pm$ according to Eq.~\eqref{hamilt}. The problem hence corresponds to the RFIM at coupling $J/2$ and field strength
$\Delta/\sqrt{2}$. In this case, the task of finding ground states is
equivalent to finding a minimum $(s,t)$ cut that partitions the graph into two
disjoint sets of nodes: one that has spins down (including $s$) and one with spins up
(including $t$) \cite{dauriac:85,hartmann:book}. Here, $s$ and $t$ are ghost
vertices relating to positive and negative random magnetic fields, respectively.
Such minimum cuts can be found in a time polynomial in the number of sites based on
the min-cut/max-flow correspondence \cite{gibbons:book}, by using algorithms such as
Ford-Fulkerson or push-relabel for the flow problem \cite{kolmogorov:04a}.

For $q>2$, on the other hand, the problem of finding ground states is {\em NP\/} hard
\cite{boykov:01}. Nevertheless, a graph-cut approach for fast approximate energy
minimization of such energy functions, occurring in computer vision problems, was
proposed by Boykov {\em et al.\/}~\cite{boykov:01}, and later on developed into an approximate ground-state algorithm for the RFPM in Ref.~\cite{kumar:18}. The basic idea amounts to the
embedding of an Ising symmetry into the Potts model, such that exact algorithms can
be used to solve a partial problem. Two variants of this idea were proposed in
Ref.~\cite{boykov:01}, dubbed $\alpha$-$\beta$-swap and $\alpha$-expansion.  For the
$\alpha$-$\beta$-swap, two spin orientations or {\em labels\/}
$\alpha \ne \beta \in \{0,1,....,q-1\}$ are picked and all labels apart from $\alpha$
and $\beta$ are frozen; the update consists of a swap of the labels between
regions. In contrast, for $\alpha$-expansion one picks a label $\alpha$ and attempts
to expand it while freezing all the remaining labels, cycling through the labels in
turn in $q$ iterations. These methods hence correspond to downhill optimization
techniques, but with a highly non-local move set, such that many (but not all)
metastable states are avoided. In practice, we focus on the $\alpha$-expansion move
as this is found to be somewhat more efficient for our problem. For a more detailed
discussion of these minimization techniques see
Refs.~\cite{boykov:01,kolmogorov:04a,kumar:18}.

\subsection{Ground-state extrapolation}

For a fixed disorder sample $\{h_i^\alpha\}$, applying $q$ iterations of
$\alpha$-expansion provides a metastable minimum or candidate ground state. By nature
of the approach, this state also depends on the initial configuration of spins
$\{s_i^I\}$. Hence a strategy for further improving the minimization results consists
of performing repeated runs for several initial configurations and picking the run
resulting in the lowest energy. If the probability of finding the exact ground state
in one run is $P_{0}(\{h_i^\alpha\})$, the success probability for $n$ runs increases
exponentially \cite{weigel:06b,kumar:18},
\begin{equation}
  P_s(\{h_i^\alpha\})=1-[1-P_{0}(\{h_i^\alpha\})]^n,
  \label{eq:success}
\end{equation}
such that the method becomes exact in the limit $n\to\infty$. This is also evident from the following observation: if one
tries all possible $q^N$ initial conditions in this way (where $N$ is the total
number of spins), the monotonous nature of $\alpha$-expansion guarantees that (at
least) the run starting with the ground-state $\{s_i^0\}$ as an initial condition
will also end in the ground state. It is hence justified to extrapolate the relevant
disorder averages in $n$ to probe the true ground-state behavior.

As a consequence of such a procedure, we consider $n$-dependent averages of the following observables: the magnetic order parameter \cite{wu:82a}
\begin{equation}
  \label{mag}
  m(n)= \frac{q\rho-1}{q-1},
\end{equation}
where
\begin{equation}
    \rho(n)=\frac{1}{N}\max_\alpha\left(\sum_{\alpha} \delta_{s_i,\alpha}\right)
\end{equation}
is the fraction of spins in the preferred orientation; 
the bond energy
\begin{equation}
  \label{ej}
  e_J(n)= - \frac{1}{N}\sum_{\langle ij \rangle} \delta_{s_i,s_j},
\end{equation}
as well as the relative deviation from the ground-state energy $E_0$,
\begin{equation}
  \label{acc}
  \varepsilon(n)=\frac{E_0-E}{E_0}, 
\end{equation}
where $E = {\cal H}(\{s_i\})$, which we call the {\em accuracy\/} of the approximation; and, finally, the ground-state overlap,
\begin{equation}
\label{over}
 o(n) =\frac{1}{N}\sum_{i} \delta_{s_i,s_i^0},
\end{equation}
where $\{s_i^0\}$ denotes the ground-state spin configuration.

In order to evaluate $\varepsilon$ and $o$ and, more generally, to judge the quality
of approximation, it is crucial to have access to a set of samples for which ground
states are known. Such samples are, in general, hard to come by for any non-trivial
system size. Here, we make use of an alternative minimization algorithm, the
sequential tree-reweighted message passing (TRW-S) method proposed by Kolmogorov
\cite{kolmogorov:06,kolmogorov:14} which, formally, amounts to solving the dual of
the linear program defined by Eq.~(\ref{hamilt}), such that in addition to the
proposed spin configuration of decreasing minimal energy it also provides an increasing lower bound on the
ground-state energy. While the bound is normally distinct from the energy of the
proposed configuration, the proposed state must be the exact ground state in case the
two energies coincide. (Note that this is a sufficient, but not a necessary condition
for TRW-S to have found the ground state.) We ran TRW-S on many samples to select a
subset for which this condition was met and we hence
can be sure of having found the exact ground state; in the following, we refer to
these as {\em exact samples\/}. These were then used for benchmarking the technique
of multiple runs with $\alpha$-relaxation outlined above. Below, we also present numerical
results for {\it regular samples} for which the exact solutions are not known.

\section{Numerical results}
\label{result}

\subsection{Graph cuts and tree-reweighted message passing}

Let us begin by comparing the two approximation algorithms: TRW-S and the $\alpha$-expansion GC. Unlike GC, the TRW-S
method does not take into account the initial spin labeling. 
Instead, TRW-S is a probabilistic message passing algorithm where an iteration corresponds to the passing of a message for each bond. As such, it converges much more slowly to a solution than the graph-cuts approach for a single initial spin configuration (and it might require damping to even converge at all \cite{kolmogorov:06}), but the resulting individual minima are typically lower than those found by graph cuts for a single initial labelling. The power of graph cuts results from the possibility of iterating over different initial labelings according to Eq.~\eqref{eq:success}. For the TRW-S method, we hence obtain approximate ground states of improving quality on increasing the number of iterations $i$, while for GC results improve with increasing numbers $n$ of initial labelings. In order to compare their performance, we ran both techniques for the same set of 1000 distinct disorder samples, and determined the energies $E_{\min}$ of the lowest-state in $i$ iterations of TRW-S or $n$ labelings of GC, where $n$ and $i$ were chosen to result in the same CPU time $t_r$ (in seconds).

\begin{figure}
  \begin{center}
    \includegraphics[width=0.99\linewidth]{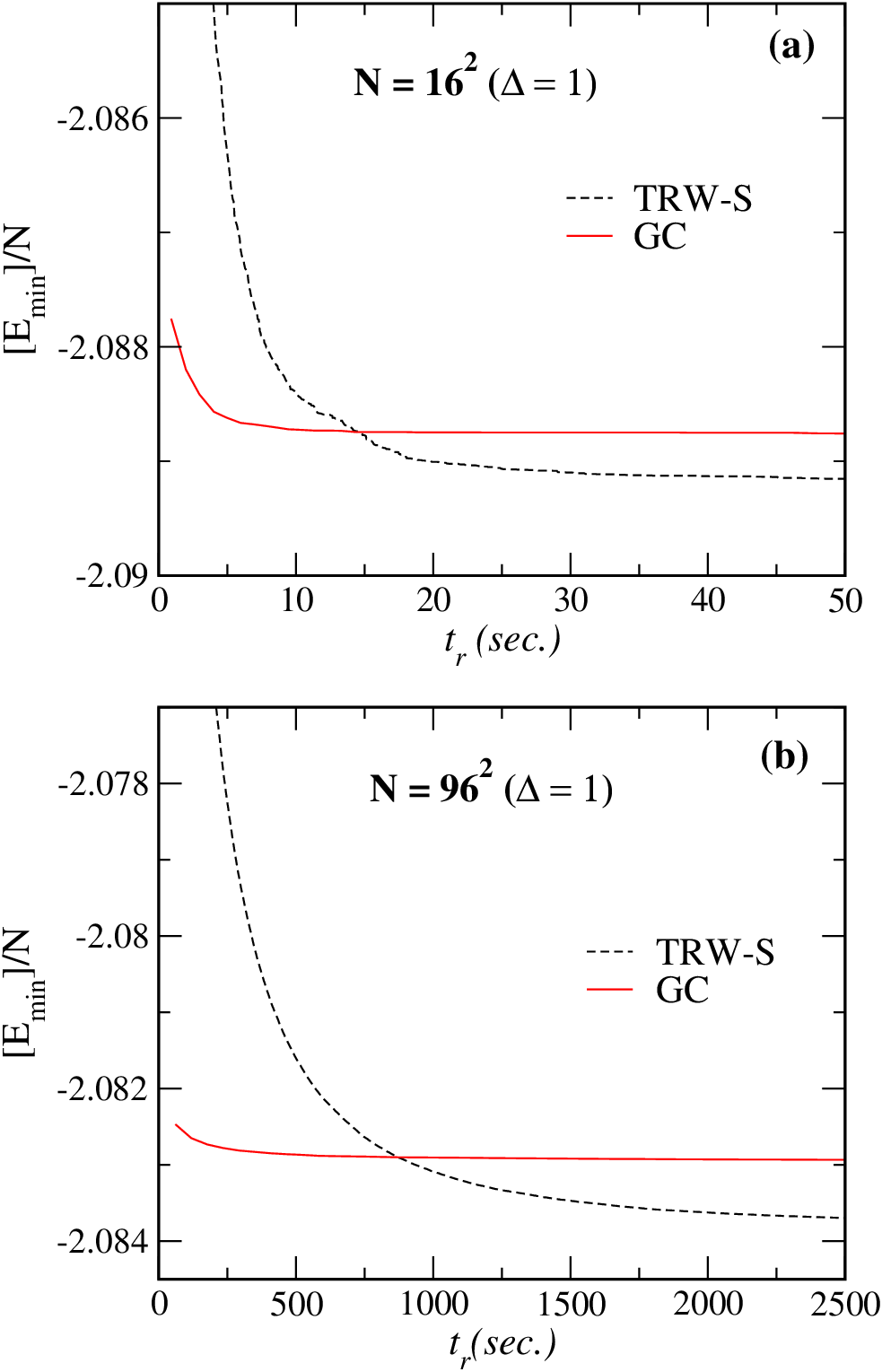}
    \caption{Plots of the disorder-averaged approximate ground-state energies, $[E_{\min}]/N$, of the $q=3$ RFPM as a function of run-time $t_r$ for TRW-S and GC on 2D square lattices (coordination number $z=4$). Panel (a) is for system size $N = 16^2$ whereas panel (b) is for $N =96^2$. The data are averaged over $1000$ random-field realizations.}
    \label{runtime_2d}
  \end{center}
\end{figure}

\begin{figure}
  \begin{center}
    \includegraphics[width=0.99\linewidth]{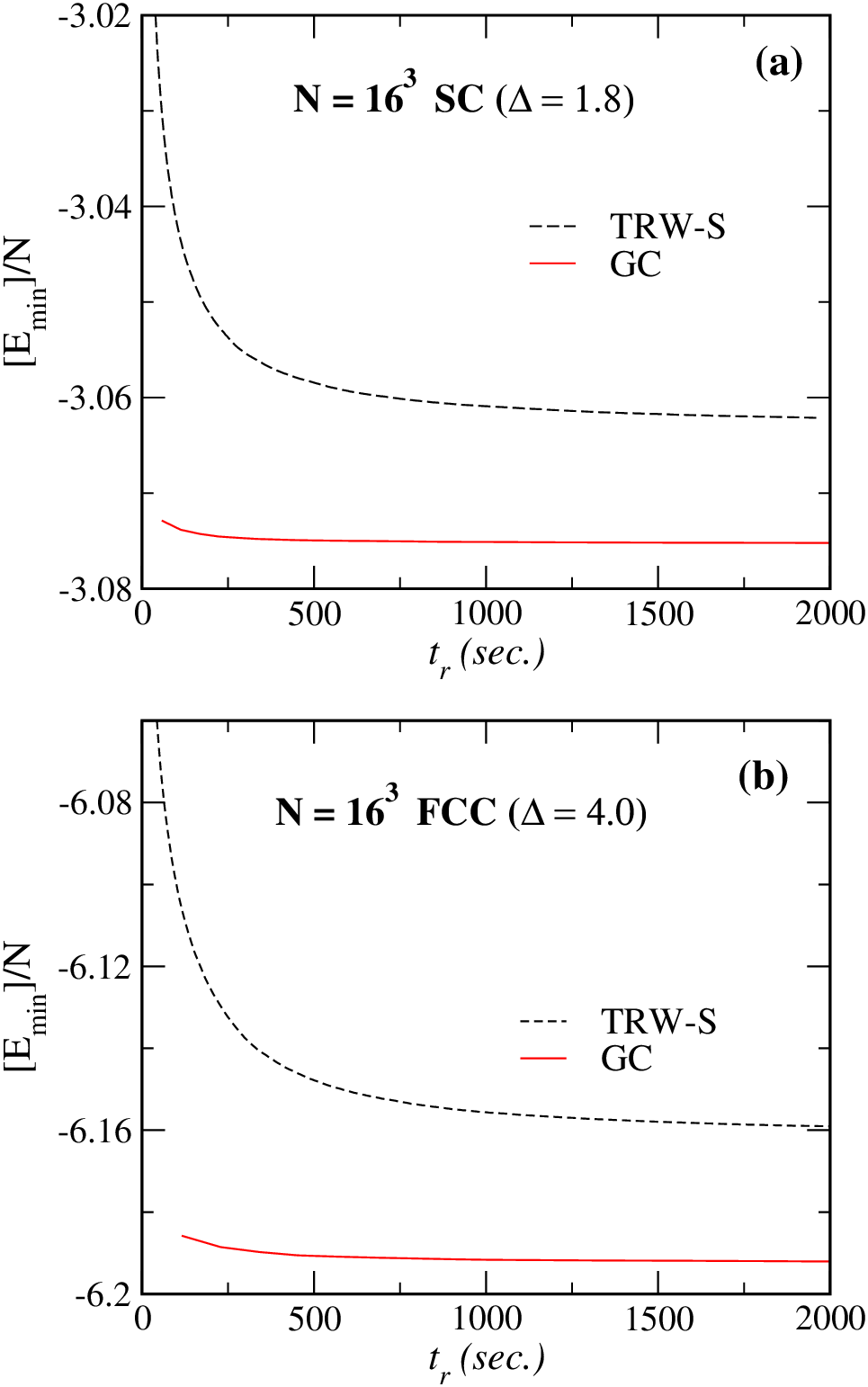}
    \caption{The same as Fig.~\ref{runtime_2d} but for systems on three-dimensional lattices. Panel (a) is for the $16^3$ RFPM system on a simple cubic lattice ($z=6$) whereas panel (b) is for the same system but on a face-centered cubic lattice ($z=12$).}
    \label{runtime_3d}
  \end{center}
\end{figure}

\begin{figure*}
  \begin{center}
    \includegraphics[width=0.95\linewidth]{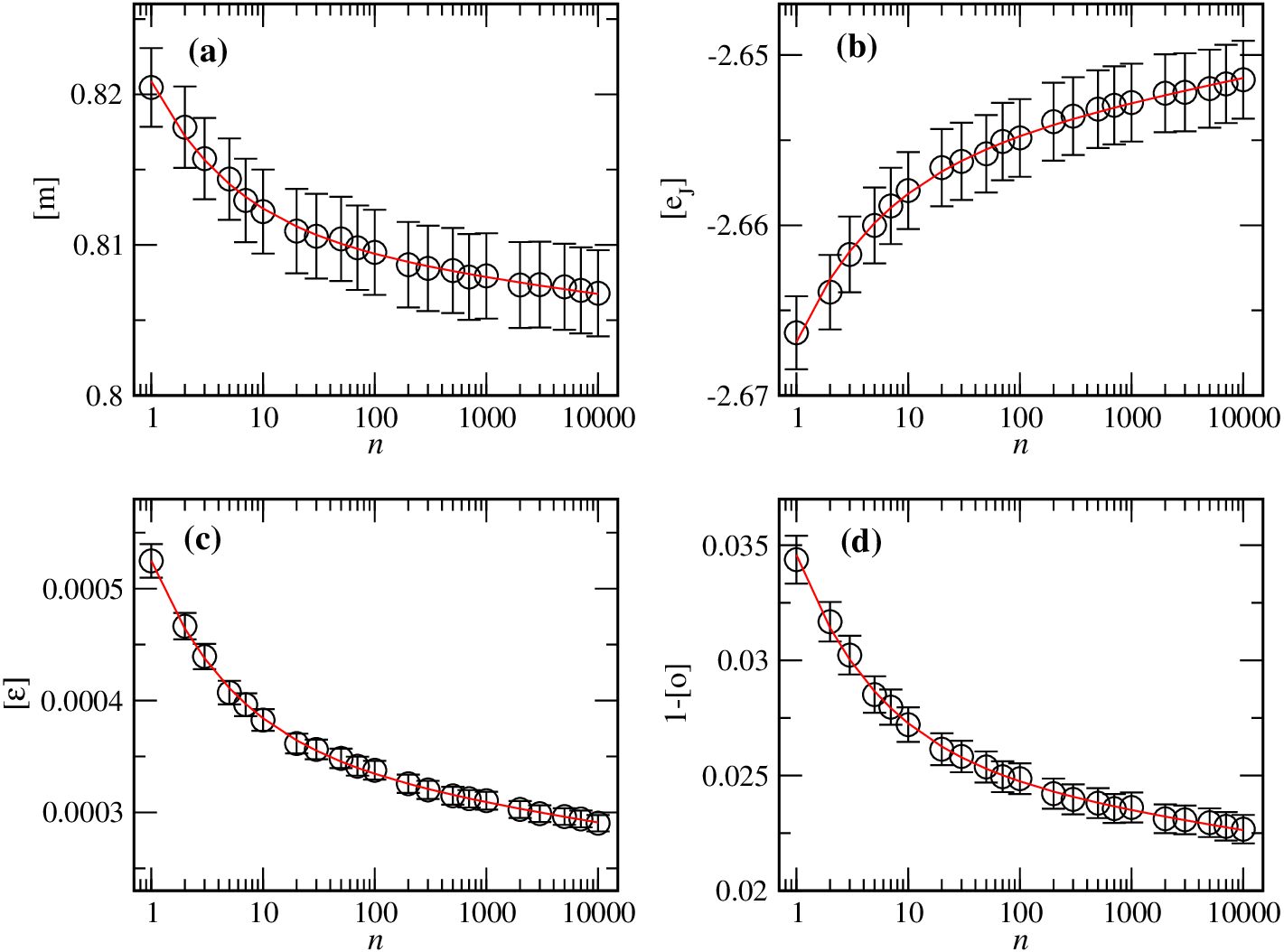}
    \caption{Disorder-averaged estimates of (a) the magnetization $[m]$, (b) the bond
      energy $[e_J]$, (c) the accuracy $[\varepsilon]$, and (d) the residual overlap
      $1-[o]$ of approximate ground states found from $\alpha$-expansion on ``exact''
      disorder samples of the $q=3$ RFPM with $N=16^3$ spins as a function of the
      number $n$ of initial conditions used.  The red lines correspond to joint fits
      of the non-linear form of Eq.~(\ref{data_fit}) to the data for all four
      observables.}
    \label{q_vs_n}
  \end{center}
\end{figure*}

Figure~\ref{runtime_2d} shows plots of disorder-averaged minimal energies $E_{\min}$ per spin, i.e., $[E_{\min}]/N$ as a function of run-time $t_r$ for samples of the two-dimensional $q=3$ RFPM on square lattices of sizes $N=16^2$ (panel a) and $N=96^2$ (panel b), respectively. The disorder samples are drawn at $\Delta = 1$, which corresponds to quite strong disorder in two dimensions \cite{kumar:18}.
It is clear from both panels that initially GC finds states of lower energy than TRW-S, but with increasing run-time there is a crossover and eventually TRW-S perform better than GC. Another observation is that in contrast to TRW-S,  GC quickly produces better approximate solutions, which then improve only slowly with the run-time.  These findings are consistent with the study of Kolmogorov~\cite{kolmogorov:06}, who compared these techniques for a stereo matching problem.

Next, in Fig.~\ref{runtime_3d}, we compare these techniques for the case of three-dimensional lattices. Panel (a) shows the comparison for a $16^3$ $q=3$ RFPM on a {\it simple-cubic} (SC) lattice for which the coordination number $z=6$, whereas panel (b) shows the comparison for the same system size on a {\it face-centered cubic} (FCC) lattice, where each spin is linked to 12 nearest neighbors via the coupling strength $J$ (i.e., $z=12$). The disorder strength $\Delta$ is chosen in both cases to be in the strong-disorder regime.  In particular, we use $\Delta=1.8$ for SC and $\Delta=4.0$ for FCC. As is clearly seen from Fig.~\ref{runtime_3d}, GC performs better than TRW-S in both of these cases. This observation is in line with previous work by Kolmogorov and Rother \cite{kolmogorov2006comparison} who compared such  techniques on vision problems for highly connected graphs. Specifically, they tested the energy minimization algorithms for stereo problems with occlusions and found that the speed of convergence of TRW-S becomes slower as the connectivity increases, and for graphs with $z > 4$ GC outperforms TRW-S. In the following, we hence focus on the use of the GC approach for our target problem of the RFPM in three dimensions.

\subsection{Extrapolation for graph cuts}

In the following, we study the RFPM for $q=3$ and $q=4$, respectively, focusing on ground state extrapolation for the simple cubic systems of size
$N=16^3$ that are large enough to provide a non-trivial benchmarking of the ground states for the
$\alpha$-expansion GC approach \cite{kumar:18}.

\subsubsection{Exact samples}

In order to generate a sample set for benchmarking, we first ran TRW-S
for $10^4$ iterations per random-field configuration and searched for exact samples for which the minimum energy $E_{\min}$ of the spin configuration becomes equal to the lower bound $E_b$ on the ground-states. For $q=3$, out of $2\times10^5$
disorder samples at $\Delta=1.8$ we found 1368 samples with exact ground states. For
$q=4$ at $\Delta=1.7$, on the other hand, 1530 out of $2\times10^6$ disorder samples
had a tight lower bound. Note that these values of the random-field strength are in
the disordered phase slightly above the transition.

We then ran the $\alpha$-expansion algorithm for these exact samples, using up to
$n_{\max}=10\,000$ different initial conditions for each random-field
configuration. From the state of lowest energy among $n$ runs, we determined the
observables defined in Eqs.~(\ref{mag})--(\ref{over}). For all
$n\le n_{\max}$, these quantities were then averaged over the total number of (exact)
disorder samples $N_{\rm samp} = 1368$ for $q=3$ and $N_{\rm samp} = 1530$ for $q=4$,
respectively. Error bars on all estimates were determined from the sample-to-sample
fluctuations.

Figure~\ref{q_vs_n} shows the disorder averaged quantities $[m]$, $[e_J]$,
$[\varepsilon]$, and $1-[o]$ of Eqs.~(\ref{mag})--(\ref{over}) as a function
of $n$ for the $q=3$ case. We find that the convergence of all averages is well
described by the same power-law form,
\begin{equation}
 \mathcal O(n)= an^{-b}(1+cn^{-d})+\mathcal O^*,
 \label{data_fit}
\end{equation}
where $\mathcal O^*$ is the asymptotic value of the quantity denoted as $\mathcal
O$. Besides the leading power law $n^{-b}$, we observe a power-law correction with exponent $d$. In the following, we present some of the evidence that justifies and explains the scaling form of Eq.~(\ref{data_fit}).

\begin{figure}
  \begin{center}
    \includegraphics[width=0.99\linewidth]{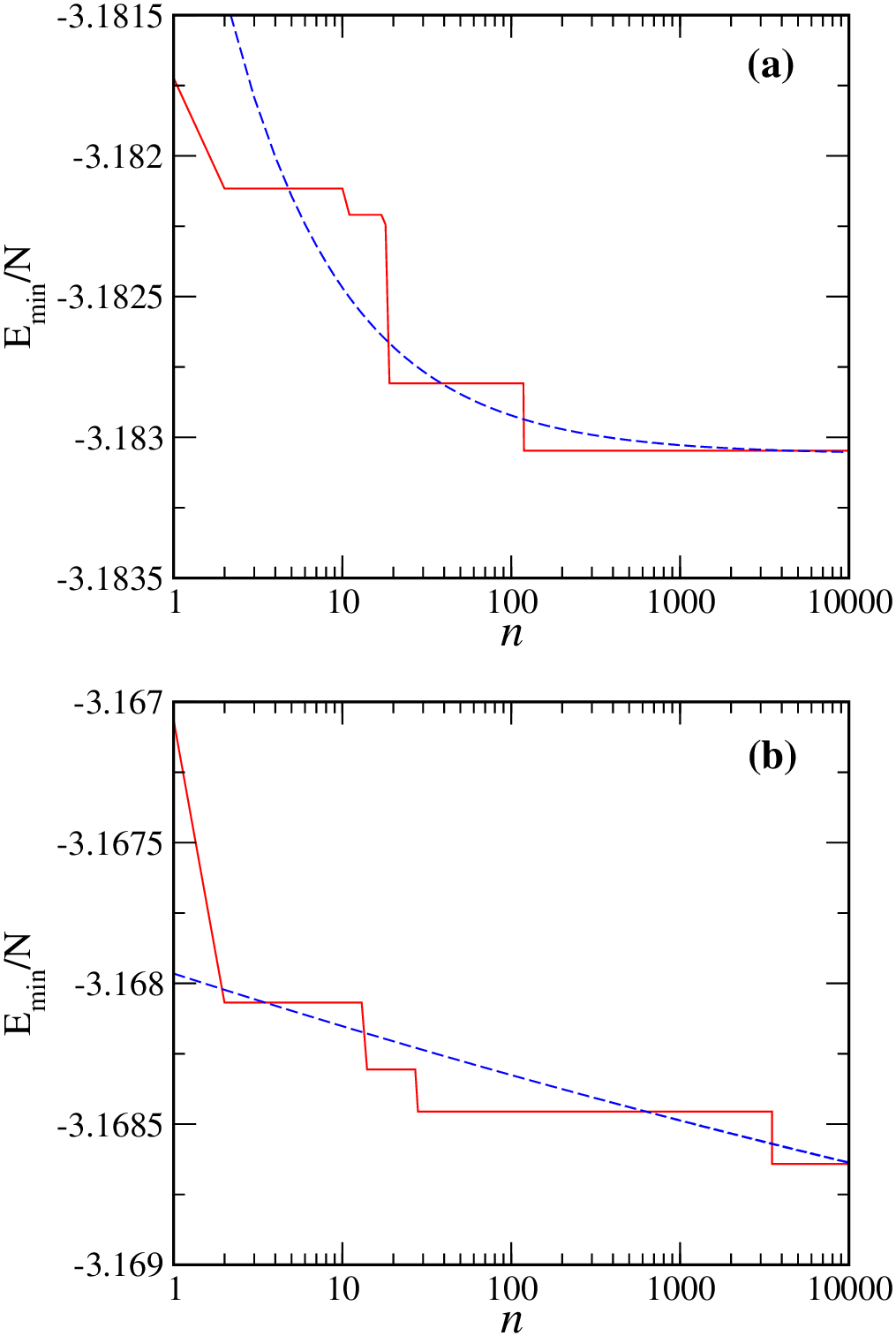}
    \caption{Plots of  $E_{\min}/N$ versus $n$ for two different samples at $\Delta=1.8$. The continuous dashed curves are fits of the form $E_{\min}=a_0 n^{-a_1}+E_0$ to the data. The values of fit parameters in panel (a) are $a_1=0.633$, $E_0=-3.18306$, and in panel (b) $a_1=0.0321$, $E_0=-3.1686$.}
    \label{Emin_gc}
  \end{center}
\end{figure}

\begin{figure}
  \begin{center}
    \includegraphics[width=0.98\linewidth]{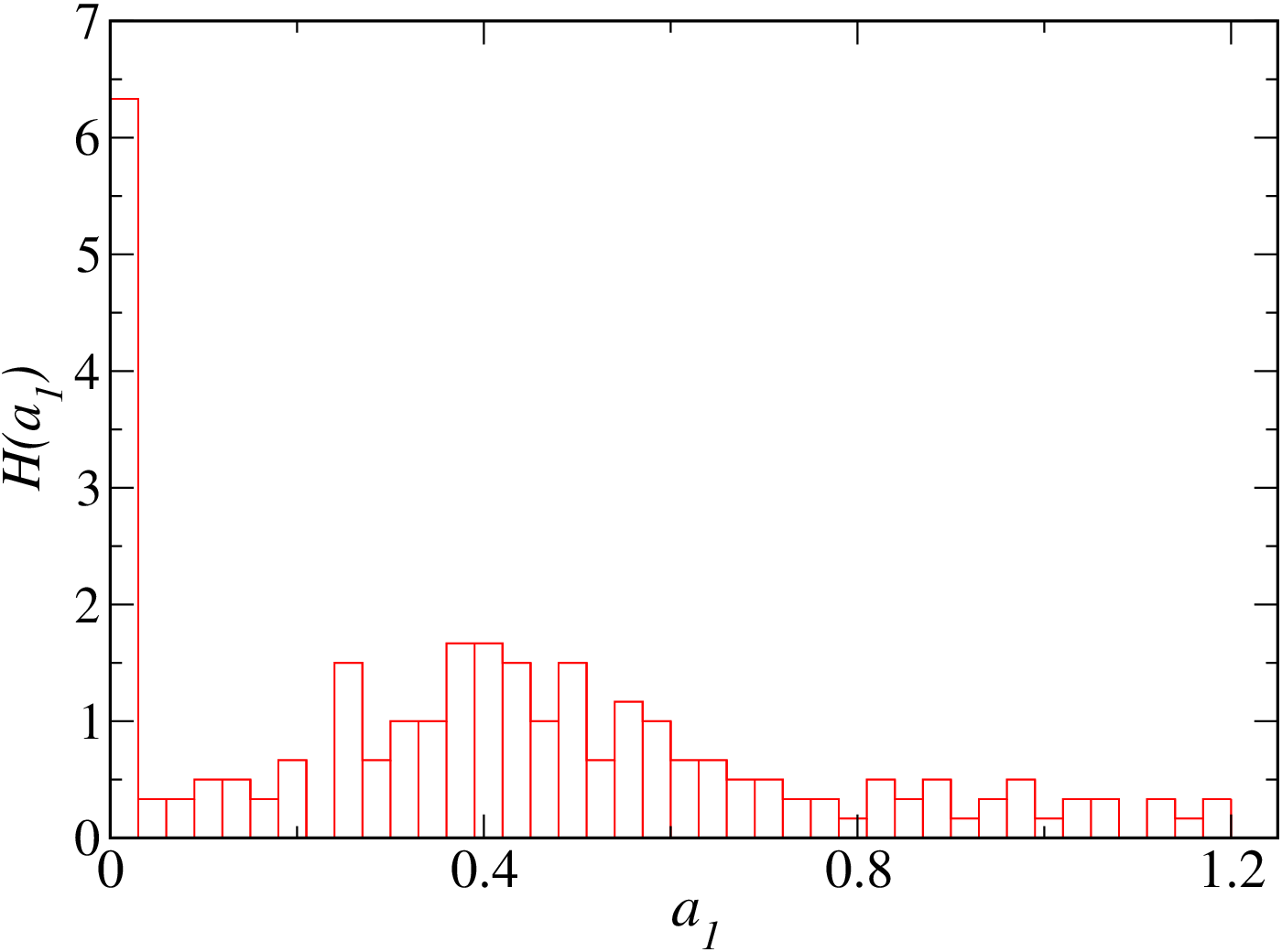}
    \caption{Histogram of 200 values of the power-law exponent $a_1$ according to the functional form $E_{\min}=a_0 n^{-a_1}+E_0$ used for fitting $E_{\min}(n)$ of individual samples according to Fig.~\ref{Emin_gc}.}
    \label{hist}
  \end{center}
\end{figure}

We analyzed the convergence  of $E_{\min}$ with respect to $n$ of $\alpha$-expansion   for 200 individual disorder samples and found two kinds of behavior: one in which  $E_{\min}$ decays markedly with $n$ and then  converges, say,  to $E_0$ for $n\to \infty$,   and the other where it  converges very slowly  in $n$. The first behavior appears due to those samples for which the approximate solutions found initially are far from the exact solutions, and hence such estimates improve considerably for lower $n$ before saturating to the exact solutions ($E_0$) or until they reach in the proximity of the $E_0$, after which they start converging slowly with increasing $n$, whereas the other behavior would be due to  those samples for which the initial approximations are near to the exact solutions and hence they show overall a slow convergence in $n$.   In Fig.~\ref{Emin_gc}, we show a typical plot of such convergences for two different kinds of samples. Notice the behavior of $E_{\min}/N$ with varying $n$, shown by the solid lines. In panel (a), $E_{\min}$ is decaying considerably already for smaller $n$ as compared to panel (b). The smooth dashed curves are fits of the form $E_{\min}=a_0 n^{-a_1}+E_0$ to the data. The value of the exponent $a_1$ in panel (a) is 0.633 and in panel (b) is 0.032. In Fig~\ref{hist} we show a histogram of 200 values of $a_1$. In this figure, the histogram clearly peaks in two different regimes; one in which the value of $a_1$ is small ($\lesssim 0.05$) and the other regime corresponds to larger value of $a_1$. Combining the two different power-law regimes containing a smaller and a larger values of $a_1$ justifies our full functional form of convergence in Eq.~(\ref{data_fit}), in which the exponent $b$ corresponds to the asymptotic exponent for slow convergence in $n$ whereas the exponent $d$ is responsible for fast decay of observable estimates for smaller values of $n$.        

Coming back to Fig.~\ref{q_vs_n},  we show the result of a joint fit of this form to
the data for all four observables, where the exponents $b$ and $d$ are constrained to share the same value among all observables, while the amplitudes $a$ and $c$ are allowed to differ for
different $\mathcal O$. The quality of fit is $Q\approx 1$. (Note that the data for
different $n$ are for the same random-field samples and hence statistically
correlated.) The resulting fits are shown together with the data in
Fig.~\ref{q_vs_n}, and it is seen that they fit the data extremely well.  The
extrapolated values of all quantities, corresponding to $\mathcal{O}^*$ in
Eq.~\eqref{data_fit}, together with the exact values $\mathcal{O}_\mathrm{ex}$ are
summarized in Table~\ref{tab}. Clearly the extrapolated and exact results are
consistent. The power-law exponents are found to be $b\simeq 0.03$ for the leading,
and $d\simeq 0.56$ for the correction exponent.

\begin{figure}
  \begin{center}
    \includegraphics[width=0.98\linewidth]{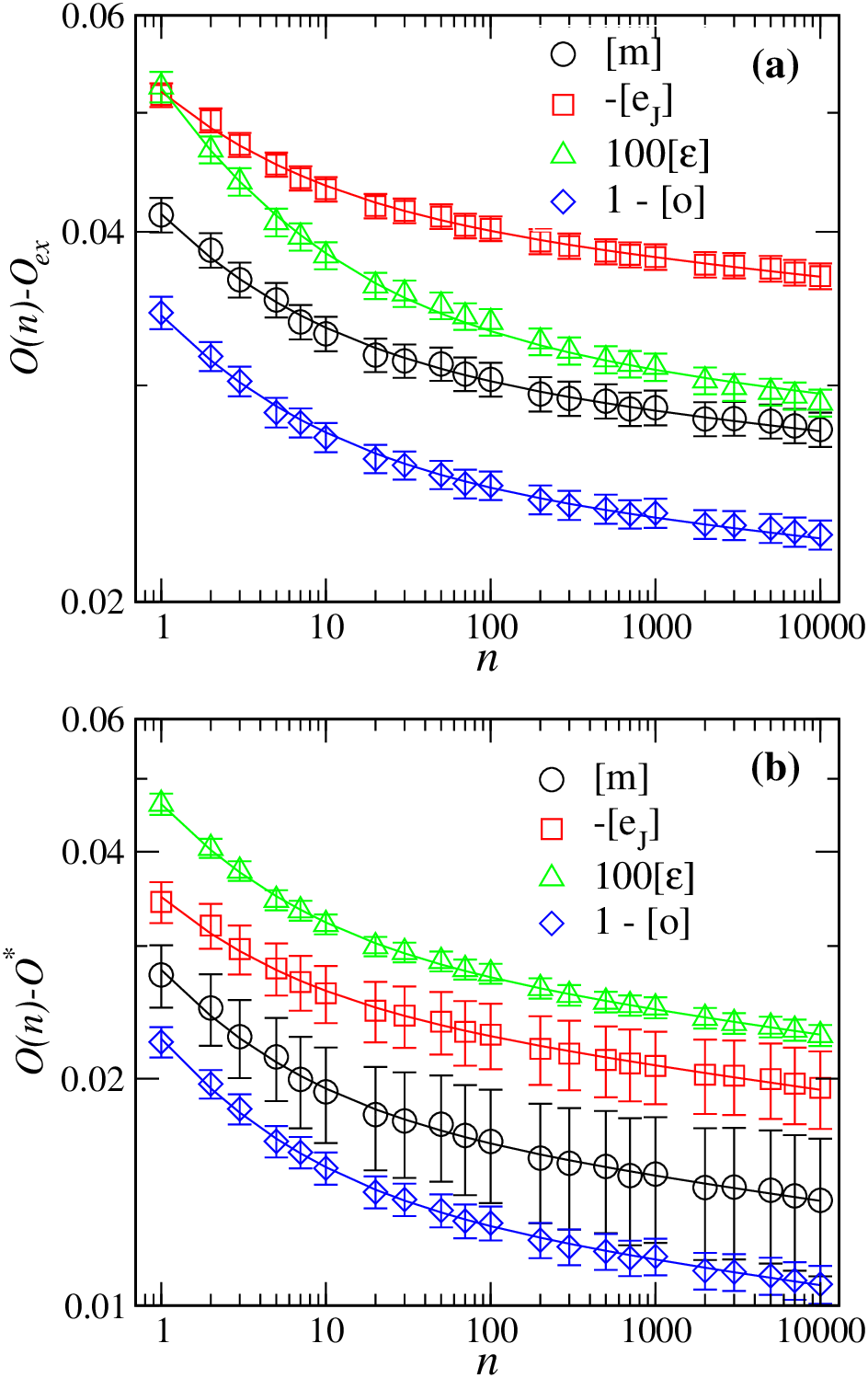}
    \caption{%
      Residuals of the observables of Eqs.~\eqref{mag}, \eqref{acc}, and \eqref{over} for the
      $q=3$ RFPM as a function of the number of initial conditions $n$ relative to
      (a) the exact results $\mathcal{O}_{\rm ex}$, and (b) the extrapolated results
      $\mathcal{O}^*$. The solid lines correspond to joint fits of the form
      \eqref{data_fit} to the data for the different observables, taking $a$, $b$,
      $c$ and $d$ as parameters.  }
    \label{residual_q3}
  \end{center}
\end{figure}

\begin{figure}
\begin{center}
\includegraphics[width=0.98\linewidth]{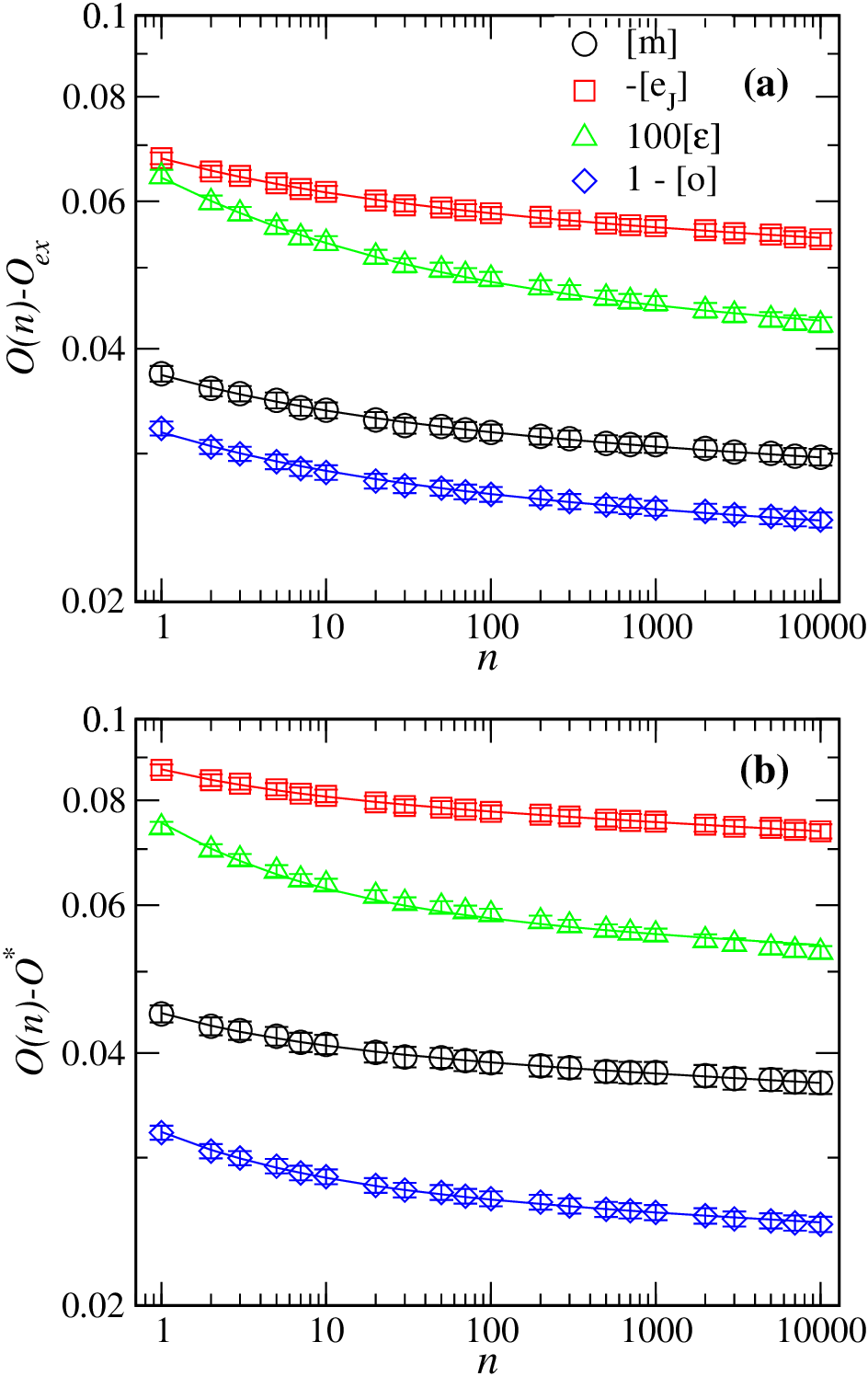}
\caption{ Analogous to Fig.~\ref{residual_q3} but for $q=4$.}
\label{residual_q4}
\end{center}
\end{figure}
 
In Fig.~\ref{residual_q3}, we plot the residuals of all quantities as a function of
$n$. Panel (a) show the residuals with respect to the exact results, i.e.,
$\mathcal O(n)-\mathcal O_{\rm ex}$, in a log-log scale. For large $n$, these decay
with $n$ in a power-law fashion $\sim an^{-b}$. However, the data for small $n$
clearly deviate from the power-law behavior, indicating the presence of scaling
corrections. This again justifies the functional form of Eq.~(\ref{data_fit}) for
describing the data, where for the values shown in Fig.~\ref{residual_q3}(b) the
limiting value $\mathcal O^*$ is taken as a constant derived from the fits shown in
Fig.~\ref{q_vs_n} and not a fit parameter. Performing a joint fit to the four
observables including all $n$ as shown by the solid curves in Fig.~\ref{residual_q3},
we arrive at the exponent values $b=0.013\pm0.005$, and $d=0.47\pm0.07$ for panel (a).  Panel (b)
shows residuals using the extrapolated results $\mathcal O^*$, determined in
Fig.~\ref{q_vs_n}. Clearly the data fit very well to the form $an^{-b}(1+cn^{-d})$,
as shown by the solid curves. The exponents $b\simeq 0.03$ and $d=0.56 \pm 0.12$ from the
extrapolated fit agree with those of the fit using the exact 
results as shown in (a).

Next, in Fig.~\ref{residual_q4}, we show the residuals for $q=4$. Again we consider
two types of residuals: (a) using the exact ground states, (b) using extrapolated
results. Also in this case, we find that the data is consistent with the behavior
$an^{-b}(1+cn^{-d})$, as shown by the solid curves. The fit in panel (a) yields the
exponent $b=0.009\pm0.005$ and $d=0.32\pm0.07$.  The extrapolated fit in panel (b)
yields the exponents $b\simeq 0.01$ and $d=0.48 \pm 0.16$, consistent with the fit
enforcing convergence to the exact result shown in panel (a). The extrapolated
estimates $\mathcal O^*$ are summarized in Table~\ref{tab} and agree with
the corresponding exact values.

\begin{table*}
   \begin{center}
%     \lineup
    \begin{tabular}{|c|c|c|c|c|c|c|c|c|}
  \hline                              
$q$&$[m]^*$&$[m]_{\rm ex}$&$[e_J]^*$&$[e_J]_{\rm ex}$&$[\varepsilon]^*$&$[\varepsilon]_{\rm ex}$&$1-[o]^*$&$1-[o]_{\rm ex}$\\ \hline \hline        
       3&0.793(14)&0.780(3)&$-2.63(2)$&$-2.615(3)$&$0.000065(72)$&0&0.012(13)&0\\ \hline 
       4&0.86(2)&0.866(2)&$-2.653(27)$&$-2.673(2)$&$0.00013(15)$&0&0.005(9)&0\\ \hline 
    \end{tabular}
  \end{center}
  \caption{ Extrapolated ($\mathcal{O}^*$) and exact
    ($\mathcal O_{\rm ex}$) results for the observable values of Eqs.~(\ref{mag})--(\ref{over}) for the exact samples for $q=3$, $\Delta = 1.8$ and $q=4$,     $\Delta = 1.7$.  The numbers in parentheses are the error estimates on the last significant figures. These error bars are calculated from fits with the  exponents of Eq.~(\ref{data_fit}) fixed to their values found from the     unconstrained fit.}
    \label{tab}
\end{table*}

\begin{figure}
\begin{center}
\includegraphics[width=0.98\linewidth]{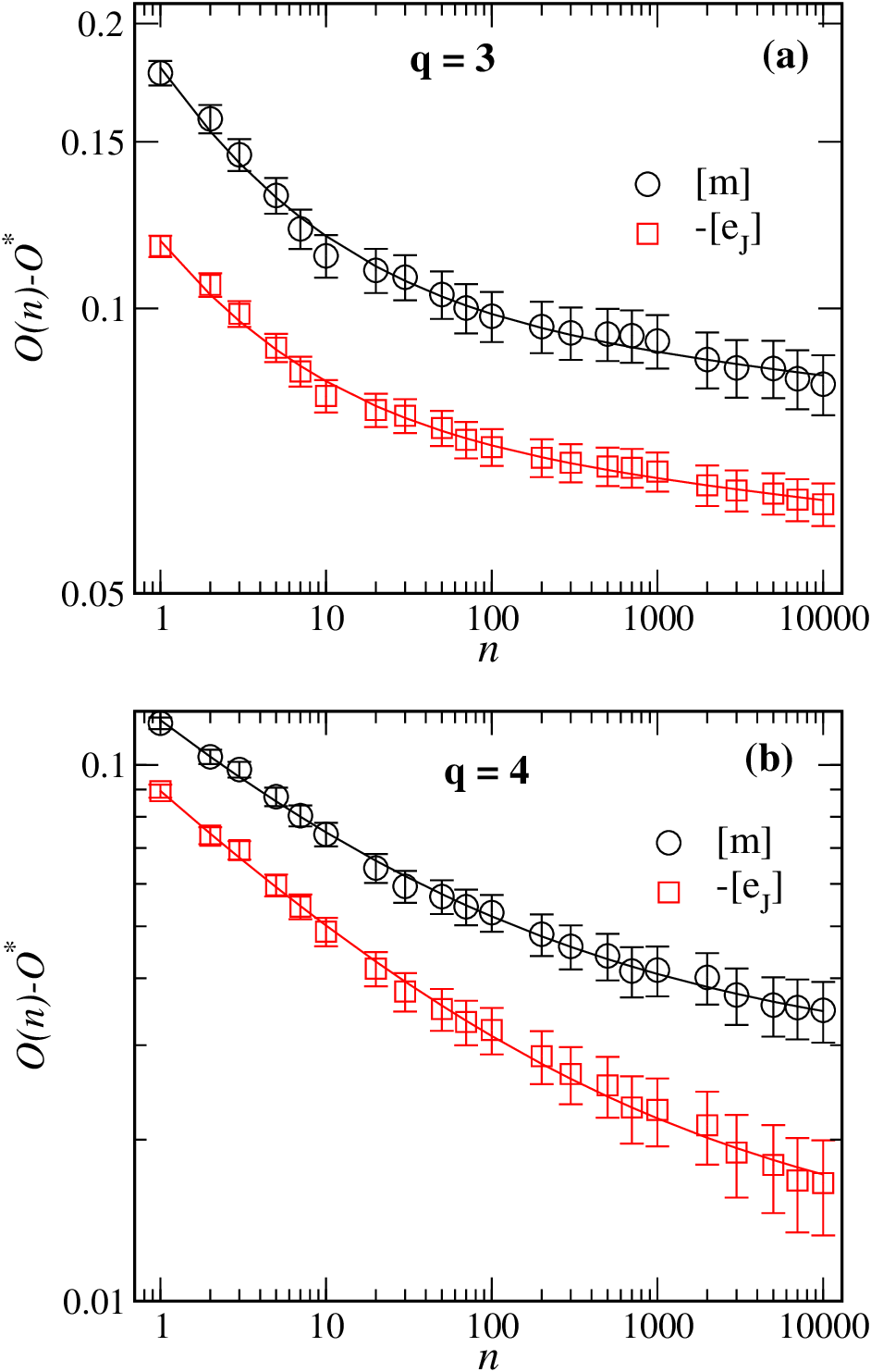}
\caption{Residual of the magnetization $m$ and bond-energy $e_J$ as a
  function of $n$ for $q=3$ [panel (a)] and for $q=4$ [panel (b)] for regular
  disorder samples
  of the RFPM.  The solid lines show joint fits of the form $an^{-b}(1+cn^{-d})$ to
  the data.}

\label{residual_reg}
\end{center}
\end{figure}

Encouraged by the observed consistency in behavior for the ensemble of exact samples,
we also considered the extrapolation behavior for the regular 
ensemble, for which
exact solutions are not available. Here we naturally cannot consider the quantities
$\varepsilon$ and $o$, and we hence focus on $m$ and $e_J$ only. For consistency, we
generated the same numbers $N_{\rm samp}=1368$ ($q=3$) and $1530$ ($q=4$) of regular
disorder samples as we had previously considered for the exact
samples. Fig.~\ref{residual_reg} shows the residuals $\mathcal{O}(n)-\mathcal{O}^*$
as a function of $n$ for both $q=3$ and $4$. We jointly fit the data of $[m]$ and
$[e_J]$ to the extrapolating form (\ref{data_fit}), shown by the solid curves. The
data fit very well for both $q=3$ and $4$ with fit-quality $Q\approx 1$. These fits
give $b\simeq 0.018$, $d= 0.51\pm 0.12$ for $q=3$ (panel a), and $b\simeq 0.0163$, $d= 0.31\pm 0.09$ for $q=4$ (panel b). These results illustrate the robustness of our theory of extrapolation in the $q$-state RFPM. The extrapolated value of observables for $q=3$ are $[m]^*=0.466\pm 0.048, [e_J]^*=-2.418\pm0.026$, and for $q=4$ are  $[m]^*=0.69\pm 0.14, [e_J]^*=-2.54\pm0.12$. 

\begin{figure}
\begin{center}
\includegraphics[width=0.98\linewidth]{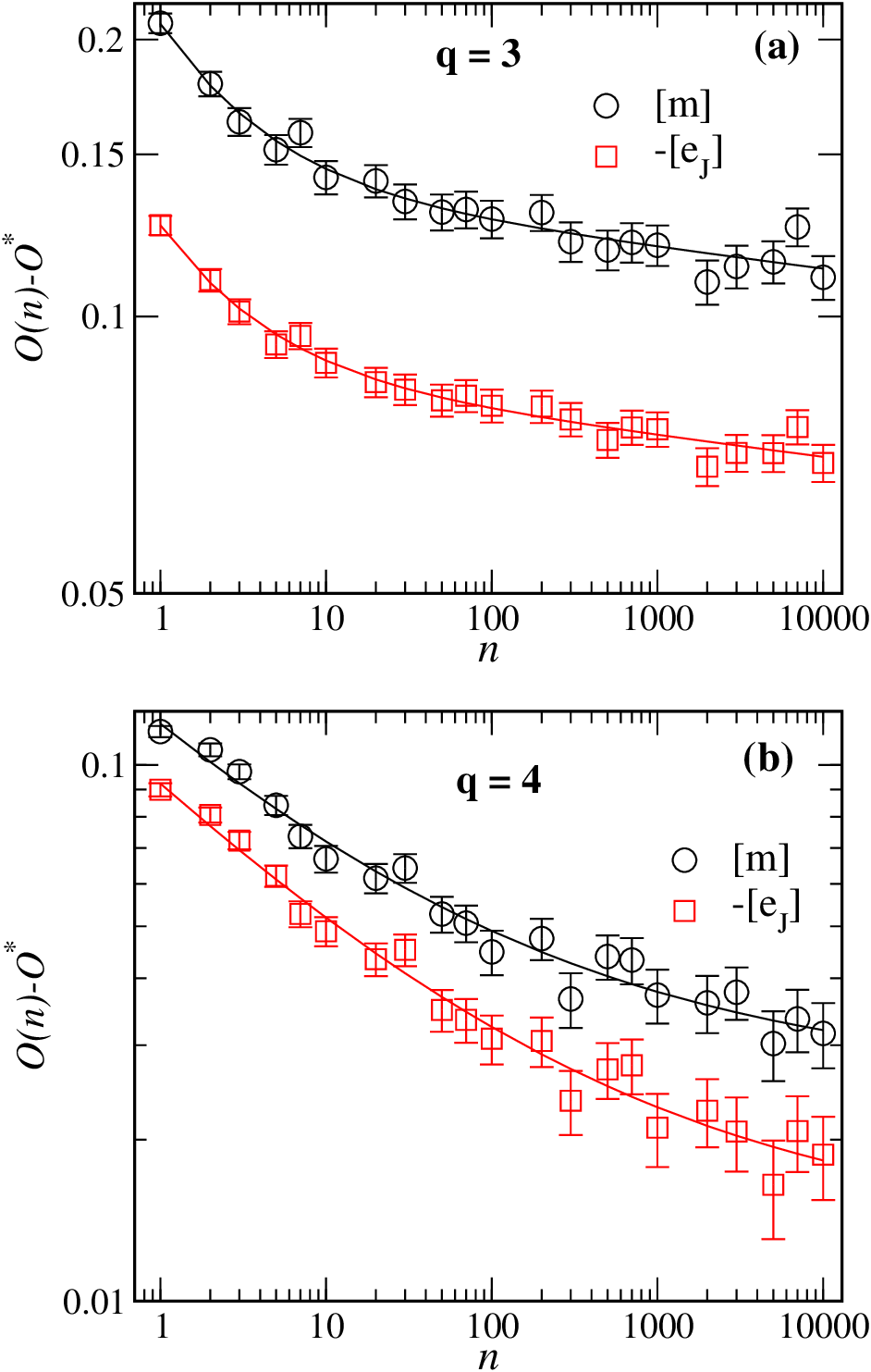}
\caption{ Analogous to Fig.~\ref{residual_reg} but the disorder samples are changed in changing the value of $n$.}
\label{uncorr_samp}
\end{center}
\end{figure}

So far the numerical results we presented are correlated in the sense that the same disorder samples are used for different values of $n$. Relaxing this assumption, in Fig.~\ref{uncorr_samp} we change the samples with varying $n$. This is shown for the case of  regular samples as we only have a limited number of exact samples. The fit results as shown by the solid lines for $q=3$ produce $b\simeq 0.023, d=0.71\pm 0.15, [m]^*=0.45 \pm 0.03, [e]^*=-2.417\pm 0.017$ with the quality of fit $Q=0.98$, and for $q=4$ give $b\simeq 0.01, d=0.311\pm 0.087, [m]^*=0.697 \pm 0.084, [e]^*=-2.536\pm 0.066$ with the quality of fit $Q=0.56$. Comparing these fit results with those of Fig.~\ref{residual_reg} where same samples are used for different $n$, we find that the exponents $b$ and $d$ slightly differ, but more importantly, the extrapolated observables $[m]^*$ and $[e]^*$ agree very well.

\section{Summary and discussions}
\label{conclude}

We have studied the performance of approximate ground-state algorithms based on graph
cuts for the three-dimensional random-field Potts model. Combining the
$\alpha$-expansion approach developed in computer vision \cite{boykov:01} with the
use of repeated runs for different initial spin configurations
\cite{kumar:18,weigel:06b} allows us to systematically improve the quality of
approximation and the results must ultimately converge to the exact ground states as
the number of initial conditions is increased. Using a collection of samples of size
$16^3$ for which exact ground states for $q=3$ and $q=4$ are available from the TRW-S
primal-dual optimization algorithm proposed by Kolmogorov
\cite{kolmogorov:06,kolmogorov:14} allowed us to illustrate this phenomenon
explicitly. Studying the behavior of the magnetization and bond energy as well as the
deviation from the ground-state configuration and energy, we found that these
quantities approach their exact values in a power-law fashion with an exponent that
is common between different quantities. Using joint fits and incorporating a
power-law scaling correction we found that our proposed scaling form fits the data
very well, and the asymptotic values of all quantities in the limit of $n\to\infty$
agree with the exact results both for $q=3$ as well as $q=4$, see
Table~\ref{tab}. For the case of regular samples that are most relevant for the
practical task of extrapolating results of the $\alpha$-expansion approach for larger
systems we find a behavior very similar to that for the exact samples, thereby
providing confidence that the extrapolation procedure outlined here will lead to
reliable results for studying the critical behavior of the random-field Potts model
more generally \cite{kumar:19}.

\acknowledgements

The authors acknowledge support by the Royal Society - SERB Newton International
fellowship (NIF$\backslash$R1$\backslash$180386). We acknowledge the provision of
computing time on the parallel compute cluster Zeus of Coventry University.

% \bibliographystyle{iopart-num}
%\bibliography{ref.bib}

%merlin.mbs apsrev4-1.bst 2010-07-25 4.21a (PWD, AO, DPC) hacked
%Control: key (0)
%Control: author (8) initials jnrlst
%Control: editor formatted (1) identically to author
%Control: production of article title (-1) disabled
%Control: page (0) single
%Control: year (1) truncated
%Control: production of eprint (0) enabled
%

\end{document}